\documentclass[twocolumn,times,nolinenumbers]{aastex631}
\usepackage{url}
\usepackage{graphicx} 
\usepackage{amsmath}
\usepackage{soul}
\usepackage{upgreek}

\newcommand{\dod}{FRB 20121102A}
\newcommand{\dic}{FRB 20190520B}
\newcommand{\ven}{FRB 20201124A}

\newcommand{\ventidu}{FRB 20220912A}
\newcommand{\be}{\begin{equation}}
\newcommand{\ee}{\end{equation}}
\definecolor{rgreen}{RGB}{0, 128, 0}
\definecolor{sblue}{RGB}{0, 56, 168}

\defcitealias{dallosso2024}{Paper~I}
\newcommand{\paperone}{\mbox{\citetalias{dallosso2024}}}

\shorttitle{Gravitationally-lensed FRBs in NS magnetospheres II: Applications}
\shortauthors{La Placa et al.}
\graphicspath{{./}{figures/}}

\begin{document}

\title{GRAVITATIONAL SELF-LENSING OF FAST RADIO BURSTS IN NEUTRON STAR MAGNETOSPHERES: \\II. APPLICATIONS TO STRONG REPEATERS AND THE CHIME POPULATION}

\author[0000-0003-2810-2394]{Riccardo La Placa}
\affiliation{INAF -- Osservatorio Astronomico di Roma, via Frascati 33, I-00078, Monte Porzio Catone, Italy} \correspondingauthor  {\url{riccardo.laplaca@inaf.it}}

\author[0000-0003-4366-8265]{Simone Dall'Osso}
\affiliation{Dipartimento di Fisica e Astronomia ``Augusto Righi'',  Universit\`{a} di Bologna, via P. Gobetti 93/2, 40129, Bologna, Italy}
\affiliation{INAF -- Istituto di Radioastronomia, via P. Gobetti 101, 40129, Bologna, Italy}
\author[0000-0002-0018-1687]{Luigi Stella}
\affiliation{INAF -- Osservatorio Astronomico di Roma, via Frascati 33, I-00078, Monte Porzio Catone, Italy}

\author[0000-0001-5902-3731]{Andrea Possenti}
\affiliation{INAF -- Osservatorio Astronomico di Cagliari, Via della Scienza 5, 09047 Selargius, Italy}



\begin{abstract}
Paper I in this series introduced a model in which seed radio bursts produced by a hotspot anchored in the magnetosphere of a highly-magnetic neutron star (NS) are greatly amplified by strong gravitational self-lensing and thus give rise to Fast Radio Bursts (FRBs). Key features of the FRB population 
naturally arise in the model from the dependence of the amplification on the relative orientation of the hot spot with respect to the rotation axis and the line of sight. Here we 
compare the model predictions with Five-hundred-meter Aperture Spherical radio Telescope (FAST) data from repeaters and with the general population of FRBs. We find that the burst energy distribution from \dod\ can be explained by assuming two antipodal hotspots in the NS magnetosphere, both producing seed bursts with the same log-normal energy distribution. This scenario implies a well-aligned system geometry, with 
the rotation axis, line of sight, and hotspot sites separated by $\lesssim 2$\textdegree.~Similar constraints are found for \ven\ and \ventidu, and weaker ones for \dic, owing to its smaller burst sample. We also show that 
precession of the NS rotation axis can explain the time evolution of the burst energy distribution from \dod\ as well as its temporary disappearance. In application to a cosmological population of 
randomly-oriented sources the model predicts distance and fluence distributions of FRBs in good agreement with those from a completeness-selected subsample of the first CHIME/FRB catalogue, provided the energy distribution of seed bursts spans a range of ${\sim10^{35}-10^{38}}$~erg.

\end{abstract}




\section{Introduction} \label{sec:intro}
Fast radio bursts (FRBs) are millisecond-long flashes of cosmological origin, now numbering in the thousands after more than 15 years of detections. 
Their origin remains elusive and many different scenarios and models have been proposed to interpret their properties \citep[e.g.][]{Platts2019,Zhang2022z}.
The energy distribution of FRBs spans several orders of magnitude up to $\sim$$10^{41}$~erg (under the hypothesis of isotropic emission) and is characterized by a high-energy power-law tail, while the distribution toward low energies is not well constrained yet \citep{Luo2020,James2022b,Shin2023}.
In more than 90\% of cases FRBs are one-off sources, i.e., only seen to emit once from a given position \citep[see, e.g.,][]{petroff22}.
\dod\ was the first to be detected as a repeating source \citep{spitler16}, thus challenging scenarios in which FRBs arise from catastrophic events that would disrupt the source.~It has since been the most studied FRB, revealing an extraordinary activity with thousands of bursts in a few tens of days, spanning a wide range of energies between $\lesssim 10^{37}$ and $\gtrsim 7\times 10^{39}$ erg \citep{li21}, and repeating with a $\sim$$157$ day periodic pattern \citep{rajwade20}.
Observations of this source with the Five-hundred-meter Aperture Spherical radio Telescope \citep[FAST;][]{nan11} showed a bimodal burst energy distribution \citep{li21} which terminates in a power-law at high energies, as is the case for the general population of FRBs.~It~is also characterized by a log-normal low-energy component reminiscent of the single pulses of many radio pulsars and rotating radio transients (RRATs), although shifted to energies higher by more than five orders of magnitude \citep[][]{Cairns2004,Burke-Spolaor2012, Cui2017, Mickaliger2018, Shapiro-Albert2018}.~Three more outstanding repeaters, \dic, \ven, and \ventidu, were also observed by FAST, showing up to hundreds of bursts in the span of a few weeks \citep{niu22,zhangetal22,Zhang2023}. 
All four sources have been precisely located and the redshift $z$ of their host galaxies determined to be 0.193 (\dod), 0.241 (\dic), 0.098 (\ven), and 0.077 (\ventidu). The first three have been associated with a persistent, compact radio source \citep{Chatterjee2017, niu22, Ravi2022, Ravi2023}. 

\begin{figure*}[ht]%
\centering
\makebox[\textwidth][c]{
\includegraphics[width=0.9\textwidth]{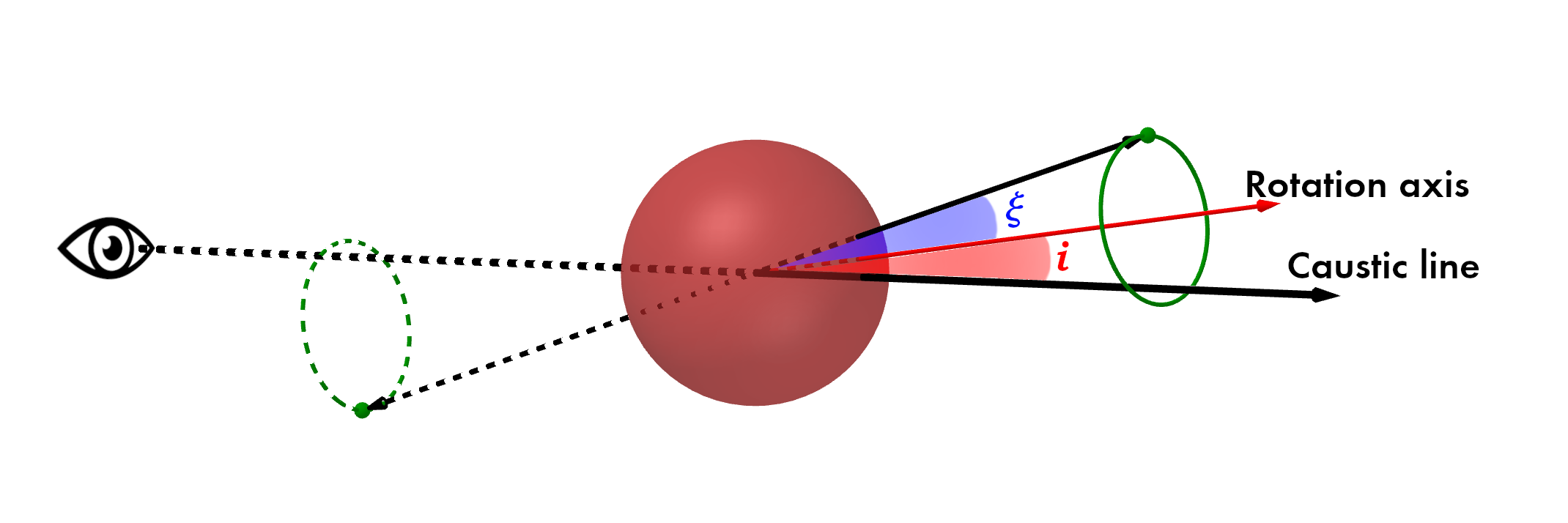}
}
\caption{Sketch of the possible geometry for \dod. The hotspots trace two circumferences during their rotation with the NS (in red), one closer to the observer (green, dashed), the other one on the opposite side of the NS (green, solid). In the further circumference, the angle from the caustic line to the rotation axis, $i$ (in red), and the emission colatitude, $\xi$ (in blue), are indicated, their size exaggerated for visual clarity.} \label{fig:1211geometry}
\end{figure*}

Many FRB models envision a cosmological population of magnetars as producing at least part of the observed FRBs \citep[see, e.g.,][]{Popov2010,Lyutikov2016,Beloborodov2017, Katz17, Margalit2018, petroff22, Zhang2022z}. The detection of FRB-like events from the galactic magnetar SGR 1935+2154, together with simultaneous X-ray emission, deeply strengthened this association \citep{chime20, bochenek20, Mereghetti2020, Lu2020, Dong22, Giri23}. 
According to a family of magnetar models, the coherent radio emission is produced inside the magnetosphere \citep[e.g.,][]{Lyutikov2016,Lyutikov2017,LyuRaf19,Lu2022,Zhu23,Beniamini2025}. 
The signal can propagate out depending on neutron star (NS) parameters, such as its rotation period and magnetic field strength, and the properties of the plasma in the magnetosphere \citep[][]{beloborodov21,qu22,Lyutikov2024}: radio emission passing through closed field lines at large distance from the star would be highly disfavoured with respect to the one moving through open lines.
A second family of magnetar-linked models invokes relativistic shocks at larger distances (e.g., \citealt{Lyubarsky2014,Metzger2019,Beloborodov2020}). However, ultra-fast shots of emission, constraining the size of the emitting region to hundreds of metres, marked polarization swings of some bursts, and the analysis of scintillation frequency scales seem to favour emission at close distances \citep[][]{Nimmo2022NatAs...6..393N,Snelders2023,Mckinven2025,Nimmo2025}.
Among possible interpretations for the periodic activity windows found so far in two sources, \dod\ and FRB 20180916B, are precession of the NS rotation axis \citep{levin20, Zanazzi2020}, or orbital precession of the system, were the NS in a binary orbit (\citealt{Chimeperiodic2020}, but see also \citealt{Zhang2022z}).

We developed a model (\citealt{dallosso2024}, hereafter \paperone) in which some key FRB properties arise from the amplification of coherent emission events in the magnetosphere of highly-magnetic NSs by strong-field gravitational self-lensing (GSL). The model envisions a hotspot, rigidly anchored to the rotating magnetosphere, which flares at random times. 
If the emission takes place when the hotspot is close to the caustic line, i.e. the continuation of the line of sight (LoS) behind the NS, its observed energy can get amplified by factors of up to $10^3 - 10^4$. 
Since the hotspot is rotating with the star, the circumference it traces depends on the inclination $i$ between the rotation axis and the caustic line, and the emission colatitude $\xi$ between the rotation axis and the hotspot position (see Fig. \ref{fig:1211geometry}): for it to pass close to the caustic line, the two angles must be 
close in value.

In that case, if both angles are large the hotspot will be close to the caustic line for a small fraction of the rotation cycle and the chances that a flare occurs at the right time to be significantly amplified will be correspondingly low.  
On the contrary, if $i$ and $\xi$ are small, the hotspot will be close to the caustic line for a much larger rotational phase interval, thus greatly increasing the probability of significant flare amplification.  
In the GSL model this leads to a straightforward interpretation, namely that the two classes of FRBs are different manifestations of a single population of active highly magnetic NSs in which $i$ and $\xi$ have close values: 
non-repeating sources are characterized by large values of the two angles and thus emit FRBs only rarely; repeating sources have small values of $i$ and $\xi$, which makes them rarer, but capable of producing FRBs much more frequently. 
When instead the two angles are significantly different, as expected in the vast majority of the sources, the hotspot flares will hardly be amplified, if at all.


In \paperone\ we 
showed that starting from seed events lensing produces a power-law high-energy tail in the observed energy distribution of flares, $dN/dE \propto E^{-\alpha}$, with $\alpha \sim 2$ in individual sources, {with possible differences at the extremes of the power law depending on the extent of the area where emission can take place}. 
This power-law tail can extend to energies orders of magnitude beyond the most powerful seed events and is quite insensitive to the shape of the seed energy distribution: the latter may be, e.g., a log-normal distribution, as with ordinary pulses of radio pulsars, or a power-law with a shallower slope, as in magnetar bursts. Amplification through lensing can alleviate the energy requirement of the brightest sources, which in the most active repeaters may otherwise deplete even the magnetic energy reservoirs of magnetars.

This paper presents a first application of the model described in \paperone\ to observations of both individual repeating sources and the general FRB population. We start by summarizing the main geometrical aspects of the model in Sect.~\ref{sec:summpaperone}. In Sect.~\ref{sec:rep} we explain the energy distribution of bursts from individual repeaters as a function of their orientation, finding it agrees well with a scenario involving two emitting antipodal hotspots, and Sect.~\ref{sec:precession} shows how incorporating precession into our model can also 
naturally 
explain both the total energy distribution of \dod\ and its 
long-term variability. 
We then extend the same scenario to a population of randomly oriented sources in Section~\ref{sec:pop}, and compare our predicted distributions with distance and fluence to the ones observed from the general population of FRBs.

\section{Summary of the GSL model geometry} \label{sec:summpaperone}

The observed intensity of light emitted in the vicinity of compact objects is affected by strong-field gravitational redshift and lensing \citep[see e.g.][]{Darwin1959,Deguchi1986a,schneider92,Mao1998,Bozza2008,Gralla2020a,Bakala2023}.
In the Schwarzschild metric, the perceived amplification in total bolometric intensity for an observer at infinity depends on three parameters: the angular separation between the source and the caustic line, $\theta$, the source's linear size, $\ell$, and its distance from the centre of the star expressed in gravitational radii, $r = R/(GM/c^2) = R/R_g$. For sources close to the caustic line, the bolometric amplification can be approximated as (\citealt{Bakala2023}; \paperone)
\begin{equation}
    \label{eq:afinitesize}
a (r, \theta) = \frac{2 \left(1 -  \displaystyle \left(2/r\right)^{0.85}\right)^{1.35}}{ \sqrt{r} \,  \theta_s \left(1+\left(\displaystyle \frac{\theta}{\theta_s}\right)^3\right)^{1/3}} \, .
\end{equation} 
where $\theta_s = \ell/R$ is the angular size of the source as seen from the centre of the NS. 
Therefore, $\theta_s$ sets the maximum value of the amplification, which is reached when the source is placed exactly on the caustic line ($\theta = 0$). At large $\theta$ values the behaviour of $a$ can still be approximated analytically (see App. A in \paperone) as it decreases monotonically, reaching values beneath one and approaching $a\rightarrow\left(1-2/r\right)^2$ for $\theta \rightarrow \pi$.

In the case of a hotspot rotating with the star, its angular separation from the caustic line is uniquely determined by the two angles $i$ and $\xi$: as it rotates, the angle $\theta$ at a given rotation phase $\varphi$ can be obtained through
\begin{equation}
\label{eq:theta}
   \cos{\theta} = \cos{\xi} \cos{i} + \sin{\xi} \sin{i} \cos{\varphi} \, ,
\end{equation}
with $\varphi =  0$ when the hotspot is at its minimum angular distance from the caustic. From Eqs.~(\ref{eq:afinitesize}) and (\ref{eq:theta}) it follows that the minimum and maximum values of the amplification, $a_{\rm min}$ and $a_{\rm max}$ are obtained for $\varphi = \pi$ and $\varphi = 0$, respectively, where $\theta = i+\xi$ and $\theta = |i-\xi|$.
The probability of obtaining a given amplification is therefore the probability of obtaining the associated $\theta$ over the circumference traced by the hotspot, that is 
\begin{equation} \label{eq:pdia-intrinsic}
   P(a) = \displaystyle \frac{\left(\mathrm{d}\theta/\mathrm{d}a\right)
   \sin \theta}{\pi \sin i \sin \xi \sin \varphi}  \, .
\end{equation}
For any fixed geometrical configuration, i.e. any combination of $i$, $\xi$, and $r$, the amplification distribution is uniquely determined and behaves as a power law, $a^{-2}$.
If the emission region is extended, either through a `jitter' in colatitude or through a radial distribution, the resulting slope becomes slightly steeper at the highest amplifications (\paperone, Sect.~4). 

{The observed energy distribution can then be obtained through the logarithmic convolution of the $P(a)$ with the assumed intrinsic energy distribution of seed events. Let us take the general case in which the bursts span an interval in intrinsic bolometric energies going from a minimum $E_{0, {\rm min}}$ to a maximum $E_{0, {\rm max}}$, with a probability density distribution $P_0(E_0)$: the resulting probability density of the observed energy distribution will be given by (see \paperone, Sect. 5 and App. B)}
\begin{equation}
\label{eq:PdEconv}
P(E) = \int_{E_{0, {\rm l}}}^{E_{0, {\rm h}}} P_0(E_0) P(a = \frac{E}{E_0}) \frac{1}{E_0} \mathrm{d}E_0  \, .
\end{equation}
{Here the integral extends over all $E_0$-values that can reach $E$ through amplification, i.e. those ranging from $E_{0, {\rm l}}~=~{\rm max} \left[E/a_{\rm max}, E_{0, {\rm min}}\right]$ to $E_{0, {\rm h}} = {\rm min} \left[E/a_{\rm min}, E_{0, {\rm max}}\right]$.}

\section{Explaining the energy distribution of strong repeaters} \label{sec:rep}


Owing to the large number of bursts they produce, repeating FRBs can place tight constraints on the energetics and short- and long-scale time behaviour of sources, providing a powerful test-bed for FRB models.
Insofar, the largest public samples of repeater bursts by a single instrument are the ones collected by FAST \citep[][]{li21,niu22,zhangetal22,Zhang2023}, the largest single-dish telescope in the world. FAST has a 90\%-completeness threshold in fluence of $\sim$0.023~Jy~ms and has accumulated hundreds of bursts in the span of weeks from a handful of repeaters.~In the following we apply our model to the FAST energy distributions from four of the best-studied repeaters, \dod, \dic, \ven, and \ventidu, with special focus on the first one.

\begin{figure}[t!]%
\centering
\makebox[\textwidth][l]{
\hspace{-1.25cm}
\epsscale{1.3}
\plotone{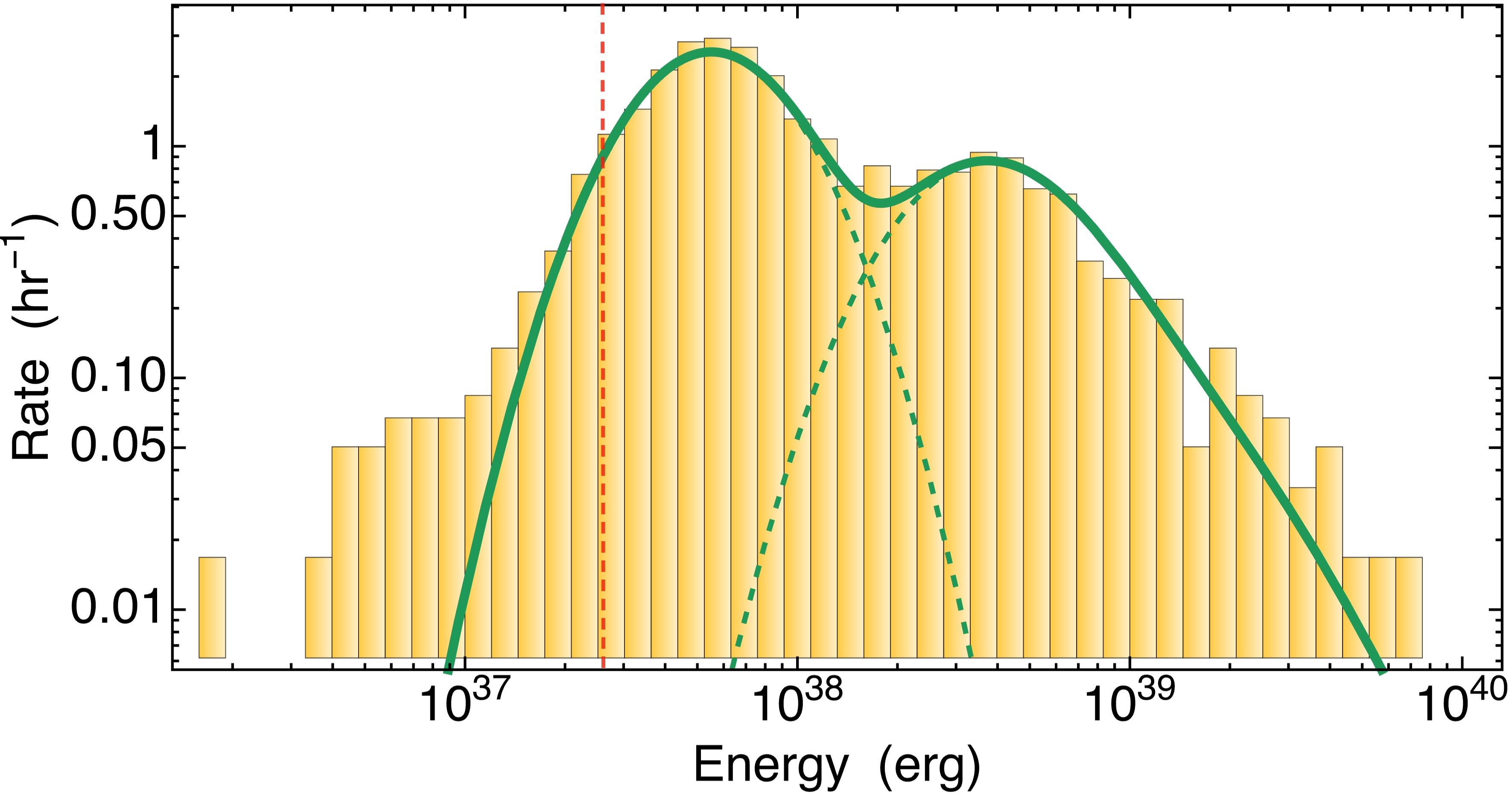}
}
\caption{The expected energy distribution of bursts from \dod\ (green solid curve) in the configuration described in Sect.~\ref{ssec:dod}, vs.~the observed energy distribution \citep[histogram;~data~from][]{li21}.~The configuration has $i=1.3$\textdegree, $\xi = 1.35$\textdegree,~plus a small jittering of the emission colatitude with semi-amplitude $\delta \xi =0.5$\textdegree.~The emitting spots have linear size $\ell = 150$~m and are located at $R=50 R_g$.~A constant event rate is assumed, while the seed log-normal has a peak (mode) value of $E_0 = 5.5\times 10^{37}$ erg and $\sigma =1/2$.~The dashed red vertical line indicates the FAST 90\% completeness threshold \citep{li21}. 
}\label{fig:pde121102} 
\end{figure}

\subsection{\dod} \label{ssec:dod}
\cite{li21} analysed a sample of 1652 bursts detected between MJD 57725 and MJD 57784, with 59.5 hours of~total observing time.~The bursts showed a bimodal energy distribution (see Fig.~\ref{fig:pde121102}) which the authors found to be consistent with the sum of a log-normal function, with a peak at $\sim$$5\times 10^{37}$~erg and $\sigma \sim 1/2$, and a generalized Cauchy function at higher energies, which displays as a power-law decay tail with index $\sim1.85$ up to $E_{\rm max} \sim 7\times 10^{39}$~erg.~This bimodality~was, however, variable over~time: after MJD~58740 the number~of high-energy events rapidly declined, while the source became completely undetected beyond MJD 58776 \citep[Fig.~1 in][]{li21}, in agreement with~the activity window and the $\sim$$157$~d period estimated by \cite{rajwade20} and most recently confirmed by \cite{Bra25}.~As a result, there were only 296 events above ${3\times10^{38}}$~erg, and 1356 below.

In \paperone\ we showed that, regardless of the geometric configuration of the source, lensing of the bursts emitted by a single hotspot always introduces a higher-energy power-law tail which skews and extends their   
seed energy distribution.~An immediate implication is that no bimodality can be introduced by GSL in that case.~However, the concurrent observation of the low-energy log-normal distribution and a second separate peak leading to the high-energy tail, 
points to a scenario with two emitting hotspots, each associated to one of the two components, reminiscent of the classic pulsar model, as sketched in Fig.~\ref{fig:1211geometry}.

In particular, if we interpret the log-normal as representative of the seed energy distribution of events in each hotspot, then one of the two must be observed with 
negligible amplification:~we can then expect this region to be facing us 
most of the time (the \textit{front pole}), so that the resulting distribution is just mildly redshifted by the NS gravitational field\footnote{If all bursts originated, e.g., from a distance $R = 50~GM/c^2 = 50~R_g $, $M$ being the NS mass, their intensity would be reduced by a factor ${\sim(1-2R_g/R)^2 \simeq 0.92}$, if they were emitted exactly along our LoS.}. On the other hand, the clear separation of the second peak from the seed log-normal indicates that the bursts from the other hotspot are all strongly lensed and magnified: therefore, its circumference always lies close to the caustic line, behind the NS relative to our LoS, and for this reason we will call it the \textit{back pole} in the following.

The span of the separation between the two peaks gives us a first indication of the minimum amplification undergone by the back pole:~in particular, a minimum $a_{\rm min}\gtrsim 5$ best describes the observed shape of the energy distribution across the dip in the $(1-5) \times 10^{38}$~erg energy range, while the maximum observed energy implies $a_{\rm max}\gtrsim 150$. As the minimum and maximum values reflect the closest and furthest points from the caustic line in the hotspot's rotation, these in turn impose (i) that the sum of $i$ and $\xi$ should be $\lesssim 4^\circ$, and (ii) that $|i - \xi| \lesssim 0.1$\textdegree\ for $\ell/R < 0.002$, $l$ being the linear size of the emission region and $R$ its distance from the NS centre, i.e. $\ell < 200~{\rm m}~(R/50 R_g)$ (Eq.~\ref{eq:afinitesize} and \ref{eq:theta}). {We choose $\ell = 150~{\rm m}$ to simplify calculations but any smaller value would fit equally well.}
Since any event produced outside of a cone of semi-aperture $\sim30$\textdegree\ around the caustic line will remain virtually non-amplified, the front pole need not be strictly antipodal to the back one, but for simplicity we will assume it to be so.

The expected energy distribution for this configuration is shown as the green solid curve in Fig.~\ref{fig:pde121102}, superimposed over the observed one (histogram).~The two dashed curves indicate the contributions from the front pole (slightly redshifted) and the back pole (strongly lensed), separately.
As a first approach, in order to account for the early disappearance of the high-energy component, and thus its smaller number of events, we imposed different normalizations for the two components, that is, different numbers of bursts from each of the two antipodal active regions, namely a 4:1 ratio between the bursts from the front pole with respect to the ones from the back pole.\footnote{We did not attempt model the 20 events below $10^{37}$ erg:~they possibly indicate a fainter tail in the low-energy component, which remains essentially undetermined as it lies below the completeness threshold \citep[vertical line in Fig.~\ref{fig:pde121102}; see also the discussion in][]{li21}.}  
However, the temporal evolution of the energy distribution in \dod\ and its 157-day modulation allow us to place further constraints on the system and better characterize its behaviour, as described in Sect.~\ref{sec:precession}.

Overall, the GSL model can account for the event energy distribution in \dod, if both our LoS and the emitting region lie within $\sim$$2$\textdegree\ of the rotation axis, and differ by $\sim $$0.1$\textdegree\ (see caption in Fig.~\ref{fig:pde121102}).~Such a configuration has a chance of occurrence\footnote{Assuming both angles $i$ and $\xi$ are randomly distributed.~If, e.g., physics constraints limited $\xi$ to be smaller than some $\xi_0$, the probability would change accordingly; for instance, setting $\xi_0=\pi/6$, it would be 3$\times 10^{-7}$.} $\sim 5\times 10^{-8}$, naturally implying an extremely small number of such powerful repeaters out to \dod's distance, $z \sim 0.2$.
We note a similar constraint was derived in \paperone, assuming all FRB sources are alike and that the large rate difference between \dod\ and the population average is due to a finely-tuned orientation of the former.

Our interpretation also implies that the actual energy budget of this source should be lower than the energy budget derived by taking fluences at face value (see \paperone, Sect.~6.3).
Based on our adopted log-normal seed energy distribution, which has a (redshift-corrected) expectation value of 7$\times 10^{37}$~erg, we derive a true energy budget $\sim 10^{41}$~erg for the 1652 bursts, 
and an average luminosity in FRBs of $\langle L_{\rm FRB}^{\rm (true)} \rangle \sim 4.5 \times 10^{35}$~erg~s$^{-1}$ during the 60 hr 
observation by FAST.
The former value is about 3 times lower than that estimated based only on non-amplified fluences (\citealt{li21}). 
Adopting a radio-to-bolometric correction factor $\sim$$10^4$, 
as estimated in the case of SGR 1935+2154 \citep[e.g.,][]{bochenek20}, the resulting power output would 
exhaust the magnetic energy of a magnetar in 
${\sim 10^3 (\eta/0.1)^{-1}B_{16}^{2}}$~yr, with $\eta$ the FRB duty cycle and 
$B_{16}^{2} $ the interior magnetic field in units of $10^{16}$~G.

\begin{figure*}[t!]%
\centering
\makebox[\textwidth][c]{
\hspace{-0.25cm}
\includegraphics[width=0.51\textwidth]{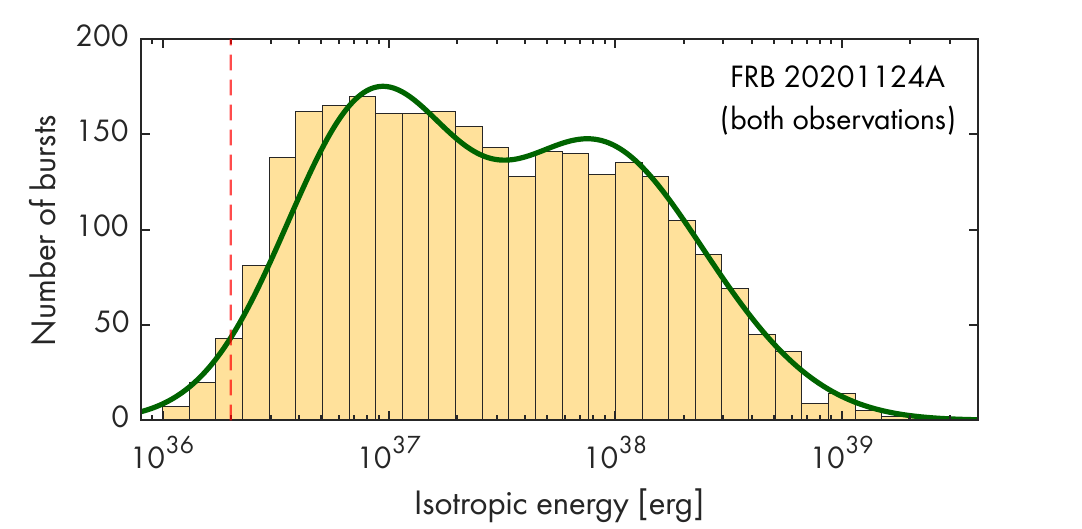}
\includegraphics[width=0.515\textwidth]{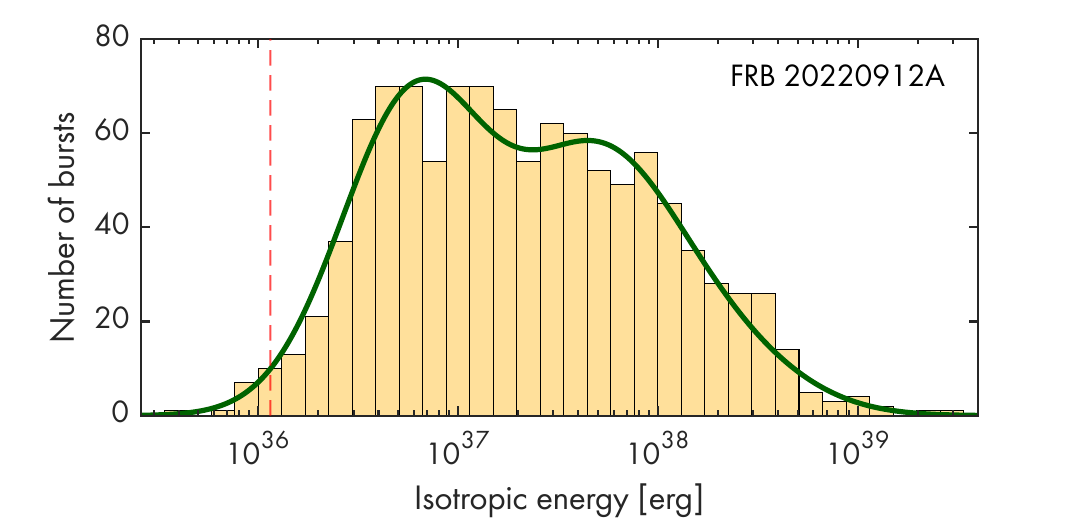}
}
\caption{\textit{Left:} Energy distribution of bursts from \ven\ (histogram; data from both 2021 observational campaigns published in \citealt{zhangetal22} and \citealt{xu22}) compared with the two emission components expected from our model as described in Sect.~\ref{ssec:ven} 
(green solid curve). \textit{Right:} Same as left panel, but for the bursts from \ventidu\ observed between MJD 59880 and 59936 \citep[data from][]{Zhang2023}; see the discussion in Sect.~\ref{ssec:2209}.}
\label{fig:20201124-20220912}
\end{figure*}

\subsection{\ven} \label{ssec:ven}

\ven\ is a repeating source first discovered in 2020 \citep{CHIME20201124A2021} by the Canadian Hydrogen Intensity Mapping Experiment \citep[CHIME][]{CHIMEFRBCollaboration2018} and located within a barred galaxy at redshift 0.098, SDSS J050803.48+260338.0 \citep[][]{Day2021,Kilpatrick2021,Nimmo2022ApJ...927L...3N}. It was the subject of two different observational campaigns by FAST: first with 91 hours of on-source time over 72 days starting in April 2021 \citep{xu22}, and then for 4 hours over 4 days in September 2021 \citep{zhangetal22}. The two campaigns yielded, respectively, 1863 and 881 bursts, with an average burst rate, in days when the source was detected, almost ten times lower in the first one than in the second. \ven\ switched off abruptly in the middle of both observation datasets 
and did not return active in the following days \citep[][]{xu22,zhangetal22}, a behaviour reminiscent of the switching off observed in \dod\ beyond MJD 58776.

~The energy distributions of the two campaigns span over the same range in energies. The second campaign shows a clear bimodality (see \citealt{zhangetal22} and Sect.~\ref{sec:precession}), while in the sum this behaviour is less pronounced (Fig.~\ref{fig:20201124-20220912}, left panel). 
A valid description of it was obtained by a model with two antipodal and identical emitting spots located at $R=50 R_g$ (see Fig.~\ref{fig:20201124-20220912}, left panel):~the seed log-normal has a median energy ${\sim 9.1 \times 10^{36}}$~erg and $\sigma \approx 0.87$, and the viewing angle and emission colatitude are respectively $i= 1^\circ$ and $\xi = 1.4^\circ$. In this case, we kept the normalization of the two components identical.
The integrated, redshift- and lensing-corrected energy of the 2744 bursts corresponds to $E_{\rm tot} \approx 3.7 \times 10^{40}$~erg, implying a factor 6 reduction with respect to the face-value isotropic energy ($\sim2.3\times 10^{41}$~erg).

\subsection{FRB 20220912A} \label{ssec:2209}

This repeater, discovered by \cite{MacandChime22} and located in the nearby galaxy PSO J347.2702+48.7066 ($z\sim 0.077$;~\citealt{Ravi2023}), was observed by FAST 
with 17 pointings, for a total of 8.67 hrs between October 28 and December 20, 2022 (\citealt{Zhang2023}). The data hint at a possible temporal variation of the energy distribution during the 50 days of monitoring.
~These observations recorded a total of 1076 bursts, sufficient to derive a reliable distribution of the isotropic energies between $10^{36}$ and $> 10^{39}$~erg. This is shown as the histogram in the right panel of Fig.~\ref{fig:20201124-20220912}.~A bimodal distribution, albeit noisy, was found to describe the data well, with two log-normal peaks at $\sim 5.3 \times 10^{36}$ and $4.1 \times 10^{37}$ erg, respectively (\citealt{Zhang2023}). Superimposed 
is the result of our model.
~The green solid curve were obtained assuming an identical seed log-normal for both emitting spots, with median $\sim 6.6 \times 10^{36}$~erg and $\sigma = 0.85$, and a geometry very close to the one previously used for \ven, viewing angle $i=1.2$\textdegree, $\xi = 1.5$\textdegree. 
The total, redshift- and lensing-corrected energy budget of the two emitting spots is $\sim 10^{41}$~erg, approximately 30 times less than the sum of the isotropic energies assumed from the observations \citep{Zhang2023}, and implying a luminosity $L_{\mathrm{FRB}} \sim 3.3 \times 10^{35}$~erg~s$^{-1}$ given the $\sim$8.7~hr of on source time.

\subsection{\dic} \label{ssec:dic}
\begin{figure}[t!]%
\centering
\makebox[\textwidth][l]{
\hspace{-0.75cm}
\includegraphics[width=0.52\textwidth]{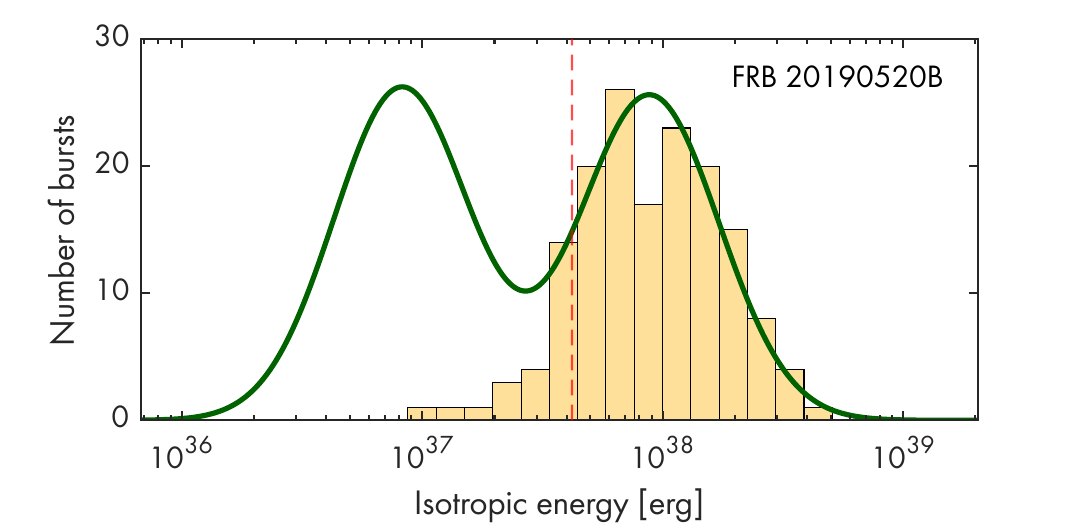}
}
\caption{The expected energy distribution of bursts from \dic\ in the configuration described in Sect.~\ref{ssec:dic} (green solid curve) versus the energy distribution of observed bursts (histogram; data published in \citealt{niu22} and \citealt{Cao25}). The 90\% completeness threshold is indicated by the red vertical line.}
\label{fig:190520}
\end{figure}

Observations of this source by FAST have detected a much smaller sample of 158 events (\citealt{niu22, Cao25}), spanning less than two orders of magnitude in $E_{\rm iso}$ and centred around $\sim 10^{38}$ erg.~Besides an intrinsically lower event rate, the single-peaked nature of the energy distribution of this source can in principle be interpreted as produced by a single visible hotspot, 
emitting at 
higher seed energies than \dod, for which amplification would give at most a small contribution to the very brightest bursts.  
However, as already discussed in \paperone\ this would contrast with the lack of a local population of such powerful sources in observations by CHIME, and would therefore require \dic\ to be a very energetic outlier with respect to the general population.
Alternatively, following the two-hotspot scenario envisioned for the other sources, a less stringent interpretation is that we only observe the second, amplified peak as a consequence of the larger source distance ($z = 0.241$; \citealt{niu22}) combined with a lower energy of the seed energy distribution, which remains completely below the FAST detection threshold.~Fig.~\ref{fig:190520} compares the bursts from \citet{Cao25} to our model by assuming two antipodal emission regions, with the same size and radial location as in \dod, and having (i) identical log-normal energy distributions, with mode at $E_{0} \simeq 5.9 \times 10^{36}$~erg and $\sigma = 0.65$ and (ii) viewing angle $i = 0.3$\textdegree~with emission colatitude $\xi = 1.5$\textdegree.~The red dashed vertical line at $\approx 4.2 \times 10^{37}$ erg corresponds to the 90\% completeness threshold of FAST observations (\citealt{niu22}), below which most events go progressively undetected.

The properties of the assumed seed log-normal imply an energy budget of approximately $3.5 \times 10^{39}$ erg for the 154 bursts and an average radio luminosity in FRBs of ${L_{\rm FRB} \sim 5.3 \times 10^{34}}$~erg~s$^{-1}$, given the 18.5~hours 
of on-source time. The overall energy budget, supposing the undetected hotspot to produce an equal number of bursts, is therefore double that, $\sim$$7 \times 10^{39}$~erg: compared to the face-value isotropic energy of all the bursts, this implies an energy reduction by a factor 5.

Note that, despite a general similarity of this source's properties with the other repeaters', 
small differences in intrinsic parameters (a narrower seed log-normal and a slightly larger $|i - \xi|$) and extrinsic ones (higher distance) can determine a remarkably different appearance of the energy distribution.

\section{Explaining evolving energy distributions and activity cycles of strong repeaters}
\label{sec:precession}

Precession of the NS rotation axis has been invoked to explain the periodic activity cycles of \dod\ and FRB 20180916B \citep{levin20, Zanazzi2020}.
%
We show here that within our model NS precession provides a natural explanation for the evolving energy distribution of the events from \dod\, which displays a decrease in time of the number of high-energy events after MJD 58740 (see Sect.~\ref{ssec:dod}),
prior to the source shutting off entirely.


As a result of precession of the NS spin axis the inclination angle $i$ changes without affecting $\xi$, thus modulating the amplification and its range.
Let us suppose to start the precession cycle from the point in which the back pole's rotation is closest to the caustic line. From there, as the spin axis precesses during the first half of the precession cycle, the circumference that the back pole traces gradually shifts away from the caustic line: therefore the maximum amplification $a_{\rm min}$ of its bursts (which is set by $|i-\xi|$) is reduced, while at the same $a_{\rm min}$ shifts towards unity as $(i+\xi)$ grows. The high-energy peak quickly shrinks in width as $a_{\rm max}$ decreases, while the ``plateau'' between the two observed components disappears as $a_{\rm min}$ approaches unity and most events from the back pole become weakly lensed, thus contributing~to the primary peak.~For instance, starting at $t_0$ with $i=1.0$\textdegree~and $\xi =1.05$\textdegree, the range of possible amplifications is $\sim (7-200)$. A precession-induced rotation of $i$ by $\sim$$10$\textdegree~at half cycle would greatly reduce the amplification range to a mere 1.4-1.8.~Given the $\sim$$157$-day precession period, the angle $i$ would be changing by $\sim$0.12\textdegree~per day, implying that $a_{\rm max} \gtrsim 100$ may be achieved for $\lesssim$~3 days around $t_0$, and that amplifications $\gtrsim$ a few tens would no longer occur after 10-15 days\footnote{Some degree of jittering in the colatitude of the hotspots, i.e. the angle~$\xi$, would allow fluctuations in $a_{\rm max}$- and $a_{\rm min}$-values among neighbouring phases of the precession cycle, which become negligible once the precession-induced change of $i$ becomes larger than the jittering amplitude.}.

\begin{figure}[t!]
\centering
\epsscale{1.25}
\hspace{-1.0cm}
\plotone{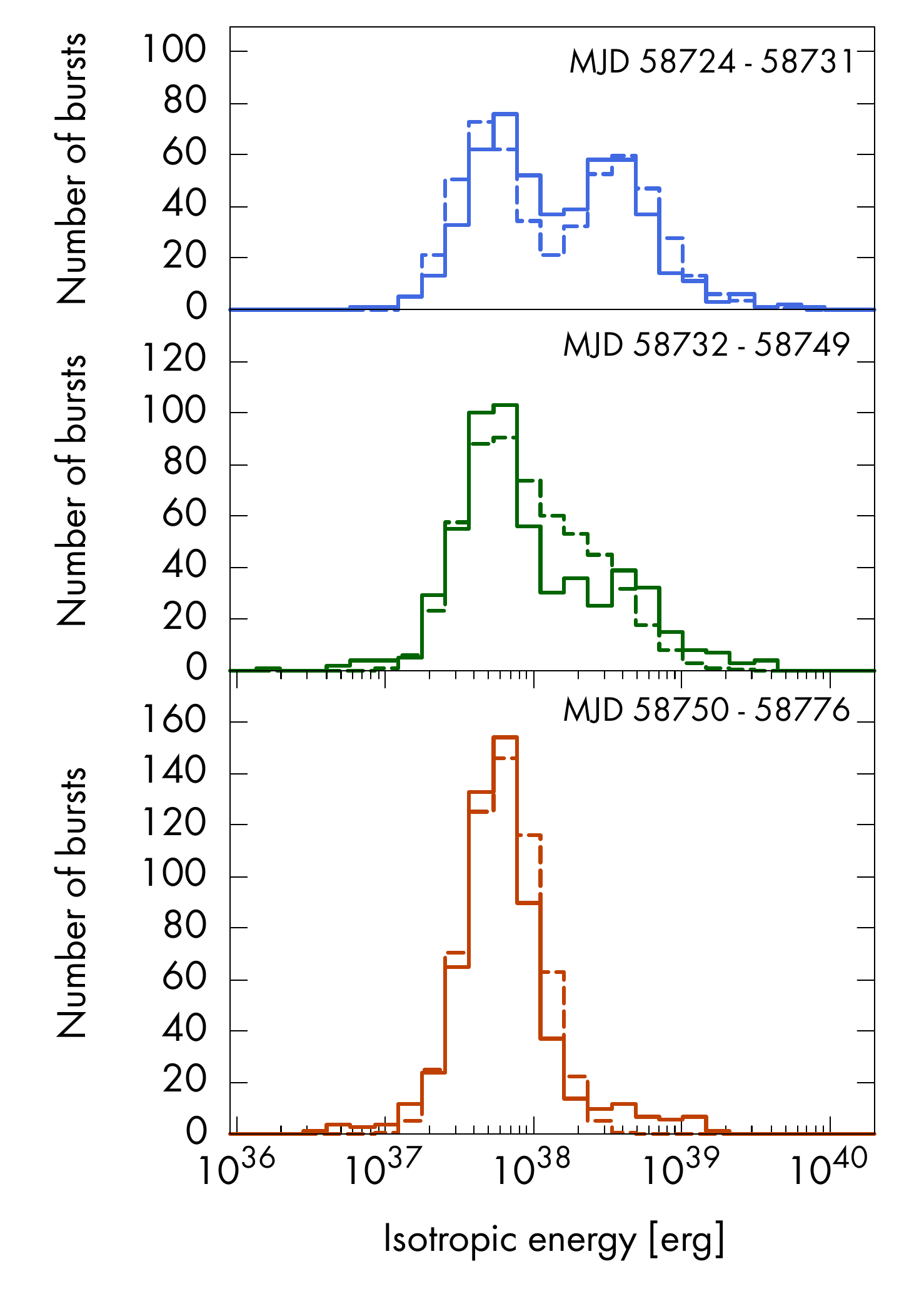}
\caption{Comparison of the observed energy distribution (solid line) of bursts from \dod\ when divided by date in three groups \citep[indicated at the top of each panel; data from][]{li21} with the expected energy distribution from our model (dashed line) integrating over the respective phases in the precession cycle.
}
\label{fig:prec_distributions}
\end{figure}

Owing to the change with time of the observed energy distribution, 
the system parameters we determined in Sect.~\ref{ssec:dod} should be regarded as approximate, average values. We carried out a more detailed comparison between model predictions and FAST observations of \dod\ in two steps: first, we 
divided the 52 days of monitoring in three different time intervals, separated at MJD 58732 and 58750 to ensure a roughly uniform and sufficiently high number of bursts in each 
interval. The 
solid lines in~Fig.~\ref{fig:prec_distributions} show~the three data groups, top to bottom, demonstrating that in the span of $\sim$7 days the high-energy peak 
diminishes with respect to the low-energy one.~In the third time interval, only one log-normal component is seen.

Then, we calculated the expected evolution of the energy distribution according to our model, following the precession cycle day-by-day, i.e.~the daily change of the angle $i$.~As before, we assumed emission by two identical, antipodal hotspots with linear size $\ell = 150$~m and located at $R = 50 R_g$, emitting bursts at the same rate, and with identical (log-normal) seed energy distributions.~Finally, we adopted $\xi=1.05$\textdegree, with jittering of semi-aperture $\delta \xi =0.5$\textdegree, a minimum viewing angle $i=1$\textdegree,~which is reached in the fourth day of monitoring (MJD 58729), and a precession semi-aperture angle of $10$\textdegree.~The variability of the intrinsic event rate was accounted for by normalizing each calculated daily energy distribution to the number of events observed in that day.~The dashed lines in Fig.~\ref{fig:prec_distributions} 
illustrate the outcome of this calculation, once daily results were grouped into the same three time windows as the FAST data.

\begin{figure}[t!]
\centering
\epsscale{1.30}
\hspace{-1.15cm}
\plotone{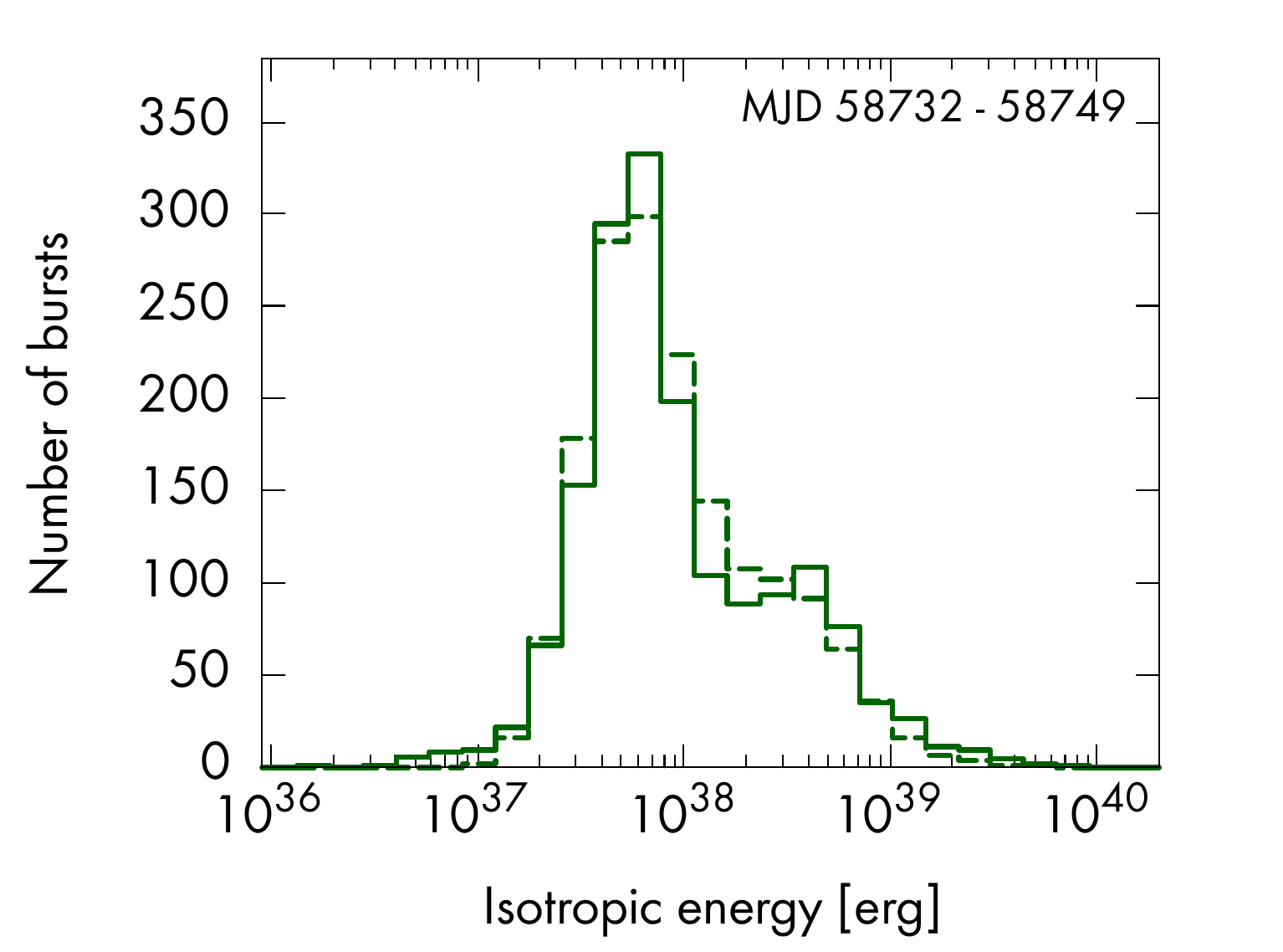}
\caption{Total observed energy distribution (solid line; same data as in Fig.~\ref{fig:pde121102}, from \citealt{li21}) compared with the expected energy distribution (dashed line; sum of the dashed lines in Fig.~\ref{fig:prec_distributions}) which accounts for the NS precession described in Sect.~\ref{sec:precession}.
}
\label{fig:prec_sum}
\end{figure}

Our result accounts well for the observed event distribution in the three different time intervals, 
and at the same time reproduces to a good approximation the overall energy distribution previously discussed: Fig.~\ref{fig:prec_sum} shows the sum of the bursts predicted by the model over the three intervals (dashed line) compared with the observed ones (solid line). 
This agreement is particularly remarkable given our simple assumption of identical and antipodal hotspots emitting the same distribution of bursts at the same rate.


Since in our scenario even non-lensed bursts are observed from this source, precession cannot directly account for its periodic shutting off.~The latter effect must be caused by some other mechanism, though coupled with the $\sim$157~d period. As we noted in \paperone\ \citep[see also][]{qu22}, absorption of coherent radio waves in the NS magnetosphere may introduce a selection on the orientation and properties of observable FRB sources. In the same vein we consider here the possibility that magnetospheric absorption, in combination with the geometry changes induced by precession, causes the periodic switch-off of \dod\ and likewise other FRBs.
In fact, absorption can be effective only along a limited range of propagation directions \citep{beloborodov21, qu22, Lyutikov2024, Xiao2024}, i.e. those sufficiently far~from the NS magnetic axis:~in this scenario, only waves travelling through open field lines at small distances from the NS may be observable \citep[cf.][]{Lyutikov2023}.~The wandering of the NS rotation axis due to precession can 
bring the emission from the hotspot in and out of the absorbing region of the magnetosphere, thus causing the windowing of the bursting activity 
of \dod\ at the $\sim$157~d period.~In this framework, the epoch at which the emission disappears corresponds to a maximum angle $\theta_{\rm max}\sim15$\textdegree, between the emission point and the line of sight (in either the back or the front pole).~A  detailed model of this effect and of its implications on the allowed source orientations is deferred to future work.

In light of the possibility that precession is driving the long-term variability in \dod, a similar assumption may be made for other repeaters. 
An interesting example is \ven: the two FAST observations analysed in Sect.~\ref{ssec:ven} are very different in length, the first one spanning 91 hours of on-source time over 72 days, while the second one only 4 hours over 4 days. 
The second dataset does displays strongly bimodal behaviour 
(see Fig.~3, bottom panel, in \citealt{zhangetal22}), while in the first, longer dataset, there is almost no dip between the two peaks \citep{xu22}. 

In the context of our model this is expected, even in absence of precession, as clear-cut bimodalities require stable geometrical configurations as the ones described above: over such a long period even small variations in either the seed distribution or the hotspots' radial and latitudinal location will always act towards a washing away of the bimodality through the sum of slightly different energy distribution shapes, which are differently weighted owing to the rate change seen from the source \citep[][]{xu22}. 
In the absence of distinct trends in the temporal evolution of the burst energy distribution however, possible long-term periodicities of the properties of the emission region, as found for \dod, could not be addressed with the available data. 
~If such temporal trends could be found in either \ven\ or other repeaters, it will be possible to apply the GSL to this variation as we did for \dod.

\section{Population study} \label{sec:pop}

In Sect.~\ref{sec:rep} we extended the simple scenario of \paperone\ to model the double-peaked energy distribution of FRBs from some of the best studied strong repeaters. 
This entailed including also the events emitted by an antipodal hotspot (in addition to the strongly lensed ones by the main hotspot) and the effects of radio wave absorption inside the magnetosphere.

As our gravitational self-lensing model envisages that both repeaters and one-off FRBs are manifestations of the same class of sources, we apply the scenario of Sect.~\ref{sec:rep} 
to the cosmic population of sources and calculate the expected redshift and fluence distributions of observed events, following the method described in \paperone.
~Both distributions are then compared to results of the first CHIME catalog \citep{chime21}.

\subsection{Model calculation}\label{ssec:thpop}

We assume that the energy of events in each source follows a log-normal distribution 
with median $\mu_i$ and width~$\sigma_i$, and that the $\mu_i$-values of different sources are also distributed according to a log-normal with median $\mu_0$ and width $\sigma_0$ 
.~The 
resulting energy distribution, $P_0(E_0)$, of all events from the assumed population will also be log-normal, with median $\mu_{\rm tot} = \mu_0$ and $\sigma_{\rm tot}^2 = \sigma^2+\sigma_0^2$.~In 
order to derive the ``lensed'' energy distribution $\hat{P}(E)$, the $P_0(E_0)$ described above must be convolved {logarithmically (see Eq.~\ref{eq:PdEconv})} with $\hat{P}(a)$, the amplification probability distribution for a population of sources containing all possible source orientations, i.e. randomly oriented viewing angles ($i$) and emission colatitudes ($\xi$). 
To calculate $\hat{P}(a)$ (and thus $\hat{P}(E)$) in the two-hotspot scenario presented in the previous section we used the prescription described in \paperone, Sect.~4.3 and 5, considering sources of linear size equal to $10 \,$m placed at a distance of $50 \, R_g$.
We also considered the effects of magnetospheric absorption of the radio pulses by introducing a viewing-angle dependent cut of the events that can reach the observer, as discussed in Sect.~\ref{sec:precession}.~For \dod\ we estimated a maximum angular separation $\theta_\mathrm{max} \sim 15$\textdegree~of the emitting region from the line of sight/caustic line, beyond which plasma effects will absorb the radio wave.~Since many different plasma and NS parameters concur to produce this effect \citep[][]{qu22, Xiao2024}, {we adopted the same value as representative of the whole population, as a full exploration of the physical parameter space would fall outside the scope of this paper.}~We then treated the cases with and without magnetospheric absorption separately. 

\begin{figure}[t!]
\centering
\epsscale{1.30}
\hspace{-1.15cm}
\plotone{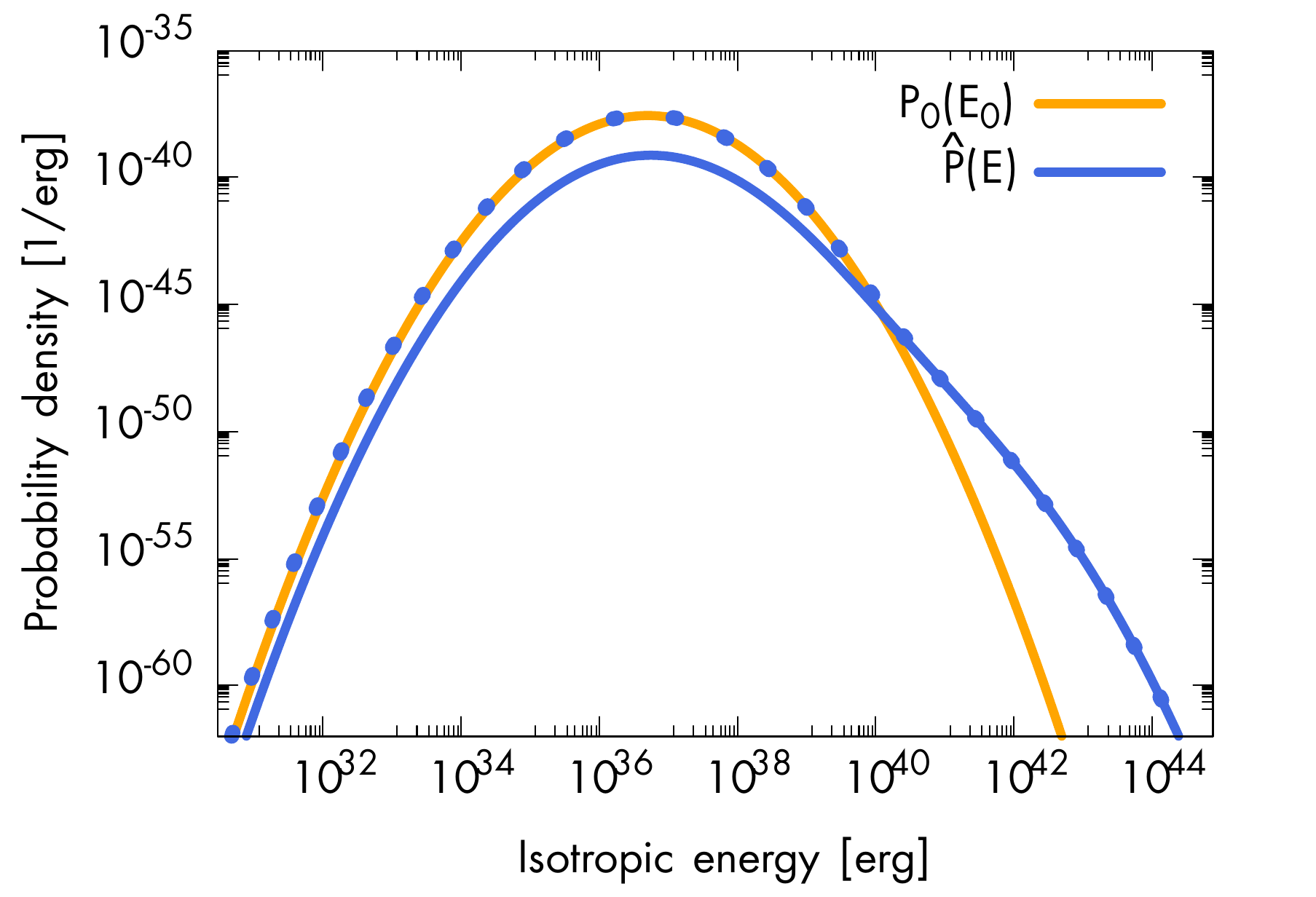}
\caption{{Example of a probability density $\hat{P}(E)$, calculated for a population of randomly oriented sources, both with and without magnetospheric absorption (respectively, solid and dotted blue lines): the magnetospheric cut is set at an angle $\theta_\mathrm{max} = 15$\textdegree. The intrinsic energy distribution $P_0(E_0)$ (orange) in this case was obtained with log-normal parameters $\sigma_{\rm tot} = 1.3$ and mode $5 \cdot 10^{36}$~erg.} }
\label{fig:pecapp}
\end{figure}

{Figure~\ref{fig:pecapp} features a sample intrinsic $P_0(E_0)$ (orange), together with the related $\hat{P}(E)$, both in the case with and without magnetospheric absorption (respectively, solid and dotted blue line): the main effect of lensing is the production of a high-energy tail extending beyond the intrinsic distribution, as already shown in \paperone, while the magnetospheric cut reduces the relative prominence of the central peak. With the present cut at $\theta_\mathrm{max} = 15$\textdegree\ the integral of $\hat{P}(E) \, \mathrm{d}E$ does not equal 1 any more, as roughly $96.6\%$ of bursts produced is hidden by the magnetosphere, and only the remaining $\sim$$3.4\%$ is observed.}
 
To compare model predictions with the observed population of FRBs 
the effect of detector sensitivity and the redshift distribution of the cosmic population of sources must be taken into account.~The former introduces a $z$-dependent minimum energy $E_{\rm thr}(z)$ in the observable $\hat{P}(E)$,
and selects an increasingly higher-energy tail of events at growing distance.
The average number of observed bursts per 
source at a given $z$ is therefore
\begin{equation}\label{eq:nbz}
N_b(z) = {\cal R}_b f_b T_b
\int_{E_{\rm thr(z)}}^{E_{\rm max}} \hat{P}(E) dE \, ,
\end{equation}
{where $f_b$ is the beaming fraction, $T_b$ the total activity time of the source, and ${\cal R}_b$ its average seed burst rate.} For the sake of simplicity we assumed $f_b = 1$ and that the product ${\cal R}_b \, T_b$ is the same for both poles and all sources. The maximum observable energy, $E_{\rm max}$, is the~maximum of the seed energy distribution $E_{0,\mathrm{max}}$ times~the maximum amplification $a_{\rm max}$, independent of redshift. 

The redshift distribution of the population was modelled assuming that FRB sources trace the cosmic star formation history, e.g. they are all young NS with no significant delay relative to the cosmic epoch of their formation \citep[][]{Wang2024}.
The NS birth rate per stellar mass $\dot{n}_{\rm NS,M}(z)$ is related to the star formation rate (SFR) $\psi(z)
$ through 
\be
\dot{n}_{\rm NS,M}(z) = K_{CC}(z) \psi(z) 
\ee
where 
$K_{CC}(z)$ is the number of NSs per stellar mass formed at a given $z$, and the SFR (\citealt{madau14})
\begin{equation}\label{eq:SFR}
\psi(z) = 0.015 \, \displaystyle\frac{(1+z)^{2.7}}{1 + \left[(1+z)/2.9\right]^{5.6}} \, \displaystyle\frac{M_{\odot}}{\rm yr~ Mpc^3} \, .
\end{equation} 
The rate of observable events within a redshift $z_{\rm max}$ is therefore (\citealt{cordes16}):
\begin{equation} \label{eq:gammabobs}
\begin{split}
& \Gamma_{b, {\rm obs}} (z_{\rm max}) = 4\pi \left(\displaystyle\frac{c}{H_0}\right)^3 \times \\
&\times \int_{0}^{z_{\rm max}} \displaystyle \frac{K_{\rm CC}(z) \psi(z) N_b(z) }{(1+z)\mathcal{E}(z)}~\left[\int_{0}^{z} \displaystyle \frac{dz'}{\mathcal{E}(z')} \right]^2 dz\, ,
\end{split}
\end{equation}
where $\mathcal{E}(z) = \sqrt{\Omega_\Lambda + \Omega_M (1+z)^3}$, and to avoid further assumptions we adopt a constant value $K_{CC} = 0.0068$~M$^{-1}_\odot$, valid for a Salpeter IMF with NS formed by progenitors in the mass range (8-40) M$_\odot$.

Taking the values for the cosmological parameters given by the 2018 Planck Collaboration results \citep[Table 2, last column in][]{planck20}, the only two unknowns in Eq.~(\ref{eq:gammabobs}) are the normalization of $N_b(z)$, \textit{i.e.}\ the value of ${\cal R}_b T_b$, and the constant needed to calibrate $E_{\rm thr}(z)$, which is set by the observational sensitivity.~To this aim we adopted the 90$\%$ completeness limit reported in the first
CHIME catalog \citep{chime21}, which corresponds to a specific fluence threshold ${\cal{F}}_{\nu}^{\rm thr} = 5$~Jy~ms for events at signal-to-noise ratio SNR~$ > 12$: we can thus translate redshifts into energy thresholds through \citep[see, e.g.][]{zhang18,James2022}
\begin{equation}\label{eq:ethrz}
E_{\rm thr}(z) = {\cal{F}}_{\nu}^{\rm thr} \, \Delta \nu \, \frac{4 \pi D^2_L(z)}{(1+z)^{2+\alpha_\nu}}  \, , 
\end{equation}
where 
$\Delta \nu$ is the detector's bandwidth, 
and $D_L(z)$ the luminosity distance, {and $\alpha_\nu$ is the spectral energy index, which we take to be equal to $-1$ following \citet{James2022c}}.

\subsection{Comparison with observations}\label{sec:compare}

For most FRB sources we do not have a measured redshift but only their dispersion measure (DM), i.e. the integrated free electron density along the line-of-sight to the source, which makes comparison with model calculations not straightforward.
The DM comprises four different contributions \citep[e.g.][]{Macquart2020}:
\begin{equation}\label{eq:DM}
\begin{split}
{\rm DM} 
         &= {\rm DM_{ISM}} + {\rm DM_{halo}} + \frac{\rm DM_{host}}{1+z}  + {\rm DM_{IGM}}(z) = \\
         &= {\rm DM_{local}}  + {\rm DM_{EG}}(z) \, .
\end{split}
\end{equation}
These arise from the Milky Way's interstellar medium (ISM) and Galactic halo, making up the local DM, and the gas in the FRB host galaxy and close environment, and from the intergalactic medium (IGM) along our line-of-sight, i.e. the extragalactic (EG) DM. 
While the first term is relatively well characterized for any source direction in the sky and the second one likely amounts to a few tens of~pc~cm$^{-3}$ \citep[][]{Keating2020}, the third term is usually unknown.
~The IGM contribution increases monotonically with $z$, as established through GRB observations \citep[e.g.][]{Ioka2003,Inoue2004,Deng2014}.~Data on limited samples of well-localized FRBs do indeed show a general monotonic growth of the total DM with redshift (albeit with some scatter) which is frequently referred to as the Macquart relation \citep[][]{Macquart2020}.~For this reason, the DM has been routinely used as a rough distance estimator for FRBs.

However, while events produced at large $z$ will inevitably show high DM, the implicit assumption that the reverse is valid 
is not always warranted: some of the FRB sources with precise localization show DM values in excess of what expected from their distance, with an extreme example in \dic\ \citep{niu22}.~In fact, it was shown that, when the measured DM is above an instrument-dependent threshold, 
observational biases make it increasingly more likely that the source is nearer than suggested by the Macquart relation, i.e.~its high value of DM is likely driven by the medium at the source, $\mathrm{DM_{host}}$, rather than by the IGM contribution \citep{James2022}.~A quantitative study of this effect in the case of CHIME was carried out in \citet{James2023} using the \textsc{zdm} code\footnote{Publicly available at \url{https://github.com/FRBs/zdm} \citep[][]{Jameszdmcode}.}:~by considering a range of possible source behaviours, the author numerically determined the function $P(z|\mathrm{DM_{EG}})$, describing the (conditional) probability with which any measured value of $\mathrm{DM_{EG}}$ can be associated to different redshifts \citep[see Sect. 7 in][]{James2023}.~After subtracting from each DM in our CHIME reference sample its ${\rm DM_{local}}$, assuming ${\rm DM_{halo}} = 50$~pc~cm$^{-3}$, we used the grid of numerical values of $P(z|\mathrm{DM_{EG}})$ calculated by the \textsc{zdm} code to convert the resulting values of $\mathrm{DM_{EG}}$ to redshifts, thus yielding a probabilistic $z$-distribution for each CHIME burst.

\subsubsection{The CHIME survey}\label{ssec:chime}

Mostly owing to its stationary transit-telescope nature, CHIME  is likely to provide the least biased survey in terms of human-driven focus on specific single sources.~For these reasons, we chose it as the benchmark against which to test our model predictions.
The first CHIME catalogue\footnote{\url{https://www.chime-frb.ca/catalog}} 
contains 536 events from 492 sources \citep{chime21}. Of these, the authors rejected for population studies all bursts with the following characteristics \citep[see Sect.~5 and 6 in][]{chime21}: 
(i) signal-to-noise ratio (SNR) $< 12$; (ii) DM $< 1.5$ times the maximum between DM$_\mathrm{NE2001}$ and DM$_\mathrm{YMW16}$, these quantities representing the Milky Way DM in the event's sky direction estimated by the models of \citet{cordes02} and \citet{yao17}, respectively; (iii) detection in the far side-lobes; (iv) scattering time above 10 ms; (v) detection on days with system concerns; (vi) complex burst morphology (in these cases, only the first sub-burst was kept).

In addition to the aforementioned cuts, when repeating sources are detected more than once, the CHIME collaboration also neglected all bursts after the first one for their population analysis. In this study, we relaxed this condition and kept all bursts from repeaters, as removing repeat bursts from the predictions calculated in Sect.~\ref{ssec:thpop} would be computationally impractical. 
We then applied a cut in fluence at the 5-Jy ms 90\%-completeness limit to the CHIME catalogue, in order to compare our model predictions to the observations. This led to the reference sample of 115 bursts which is adopted in the following, instead of the 265 used in \citet{chime21}. 
The value of ${\cal R}_b T_b$ in Eq.~(\ref{eq:nbz}) is then tuned to better follow the burst distributions in redshift and fluence from this subsample. 

{Up to shortly before the submission of this work, the largest single-instrument sample of FRB sources available was represented by the first CHIME catalogue}.  {In the meantime, the second CHIME catalogue was released \citep[][]{CHIME2026second}: it is now by far the largest collection of bursts from different sources, and will be the focus of a future comparison with a more refined simulation of sources from our model (see also Sect.~\ref{sec:summ}).}

\subsubsection{Results} \label{sec:CHIMEcomparison}

\begin{figure}[t!]
\centering
\epsscale{1.30}
\hspace{-1.1cm}
\plotone{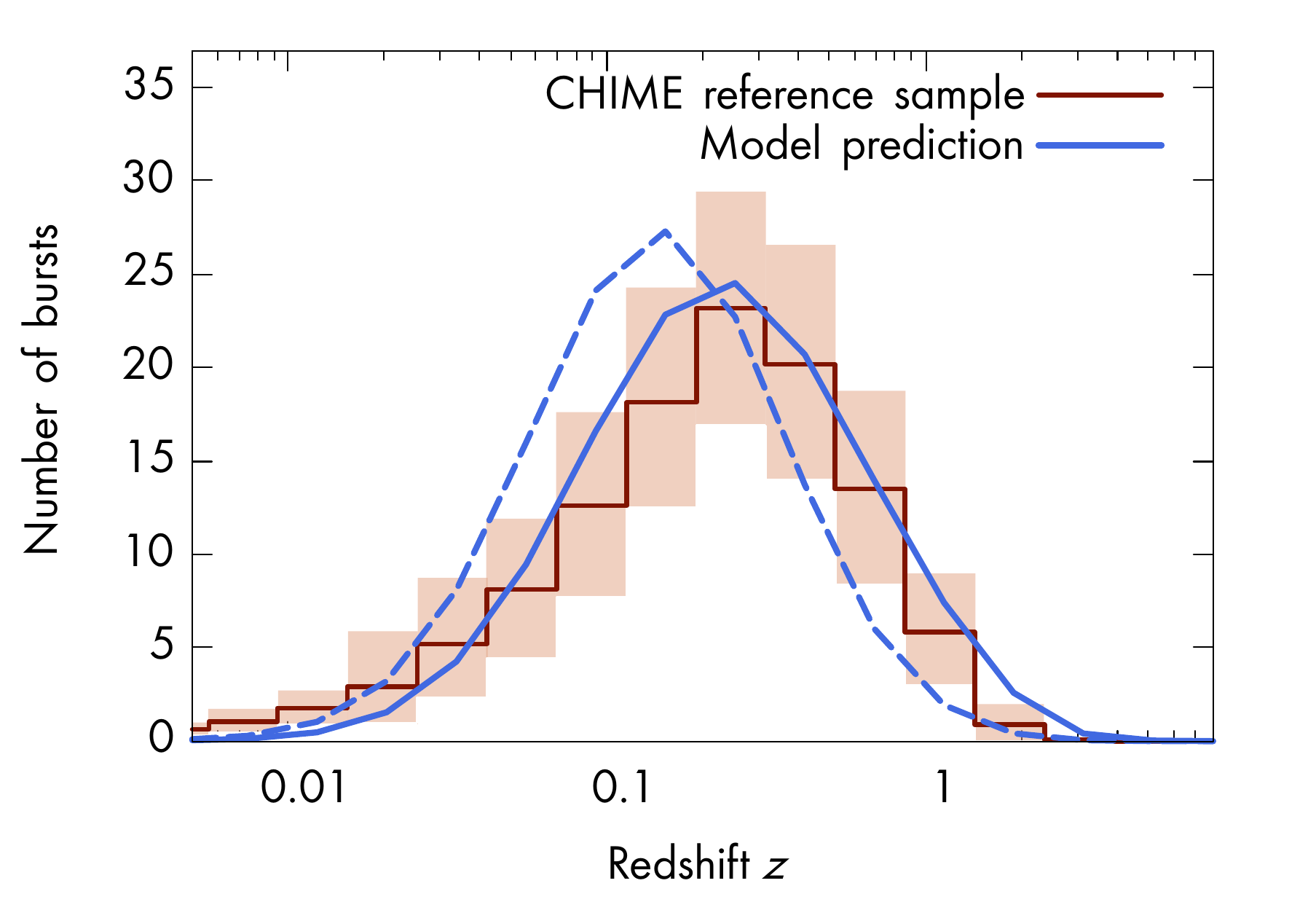}
\caption{Burst distribution with redshift: the red solid line is the average realization of possible redshifts of the CHIME reference sample, assuming a probability $P(z|\mathrm{DM})$ as in \cite{James2023}, while the shaded area represents the values comprised between the 5th and 95th percentiles. The solid blue represents the theoretical prediction {for a population with the parameters in Table~\ref{tab:simuparam}, while the dashed line corresponds to the reference case with no magnetospheric absorption (see main text).}}
\label{fig:npz}
\end{figure}

As stated in Sect.~\ref{sec:compare}, given the observed DM for each of the bursts in the first CHIME catalogue, we can obtain a probabilistic value of its redshift following the $P(z|\mathrm{DM_{EG}})$ given by the \textsc{zdm} code. Therefore, we randomly generated $10^3$ different realizations of the total redshift distribution of the reference subsample.
The average $z$ distribution thus derived for the CHIME bursts is represented in Fig.~\ref{fig:npz} by the red histogram (with the shaded area indicating the 5th and 95th percentile limits). 
{The blue solid line shows the distribution expected from a population of two-hotspot sources with a log-normal mode of $9 \cdot 10^{35}$~erg and $\sigma_{\rm tot} = 1.75$ for the $\hat{P}(E)$, and a magnetospheric cut at $\theta_{\rm max} = 15 $\textdegree.} 
{With these parameters each neutron star in the volume observable by CHIME needs only produce 58 bursts to obtain a total of almost 124 observable bursts, compared to the 115 in the reference sample \citep[taking into account the instrument field of view covering roughly 0.3\% of the sky;][]{chime21}.} 
The agreement with the reference sample distribution is quite remarkable.

The blue dashed line represents instead the expected $z$-distribution of bursts in the case with no magnetospheric absorption{: to compare the two cases, the value of ${\cal R}_b T_b$ was adjusted to produce the same total expected observable bursts. Since more bursts are visible in the case with no magnetospheric absorption, ${\cal R}_b T_b$ was lowered to slightly more than 4 to reach the same $\Gamma_{b, {\rm obs}}$.} Its poorer agreement with the data 
shows that models with some degree of magnetospheric absorption of FRBs are favoured for these parameters.
Other combinations of the parameters $\mu_{\rm tot}$ and $\sigma_{\rm tot}$ were explored, and other models with values of $e^{\mu_{\rm tot}} \lesssim 8.2 \cdot 10^{36}$~erg and $\sigma_{\rm tot} \gtrsim 1.5$ were found to be also in good agreement with the data{, with values of the resulting ${\cal R}_b T_b$ ranging from $\mathcal{O} (1)$ to the hundreds. Thus, in the case of log-normal distributions for the whole population, even supposing only magnetars to be able to produce FRBs, the number of bursts needed per source would range from tens to thousands, which is well below what a few very active repeaters have shown even in the span of a few months \citep[e.g.][]{li21,Zhang2025hyperactive}}. 

\begin{figure}[t!]
\centering
\epsscale{1.30}
\hspace{-1.1cm}
\plotone{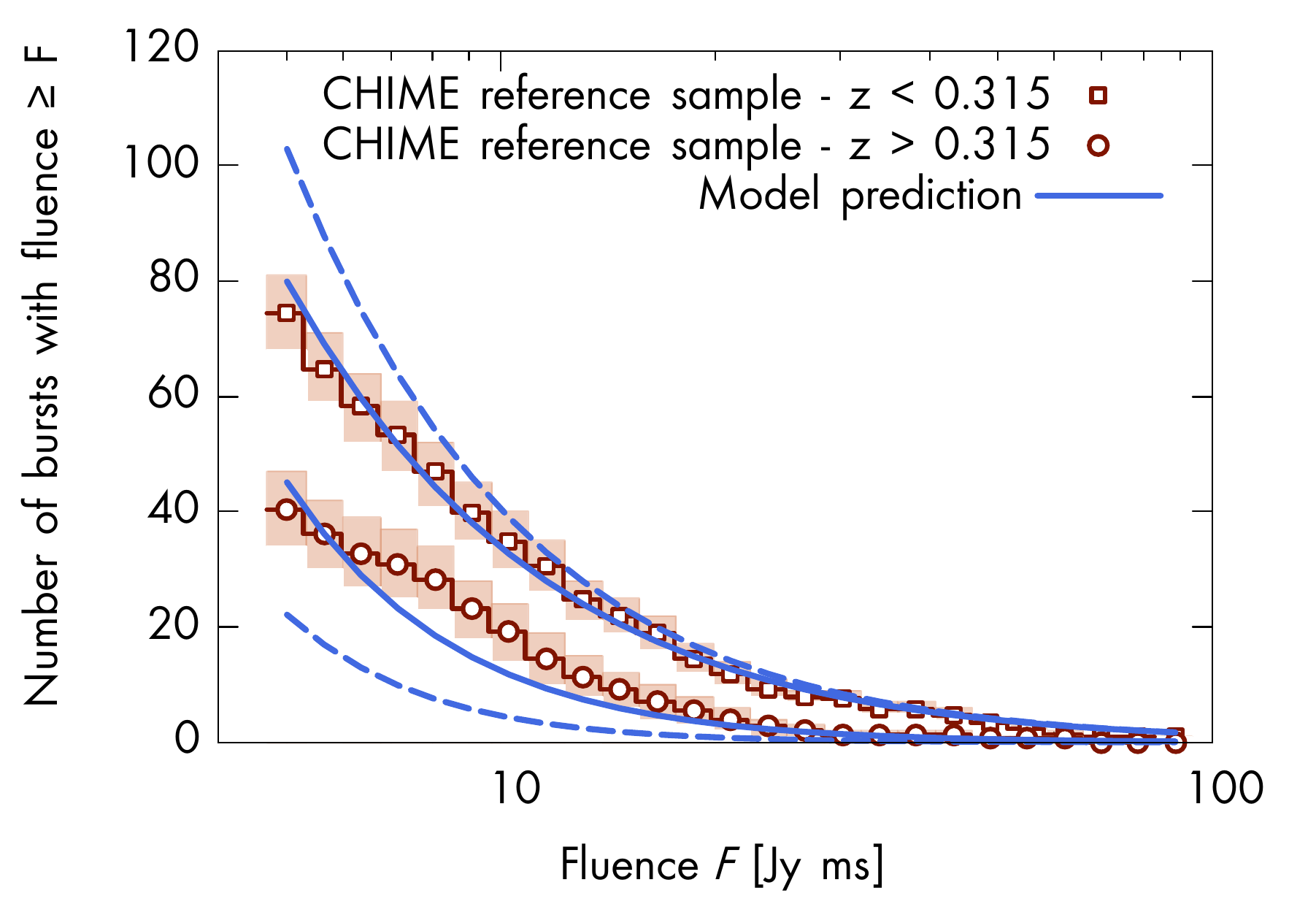}
\caption{Inverse cumulative burst fluence distribution for events below (squares) and above (circles) $z_{\rm lim} = 0.315$. The blue curves represent the expected distribution from the model, with the same parameters as in Fig.~\ref{fig:npz}, both in the case with (solid) and without (dashed) a cut at $\theta_{\rm max} = 15$\textdegree.
}
\label{fig:fludist}
\end{figure}

\begin{table}[t!]
\caption{Simulation parameters}              
\label{tab:simuparam}
\centering       
\begin{tabular}{c C}   
\hline\hline       
   Parameter                             & \mathrm{Value}         \\ 
\hline      
$\sigma_{tot}$                          & 1.75             \\
Mode ($e^{\mu_{tot} - \sigma_{tot}^2}$) & 9 \cdot 10^{35} \,\mathrm{erg} \\
Median ($e^{\mu_{tot}}$)                & 1.92 \cdot 10^{36} \,\mathrm{erg} \\
 ${\cal R}_b T_b$                       & 58           \\
$\Gamma_{b, {\rm obs}}\cdot0.3\% ^\dagger$                 & 123.94 \, \mathrm{year}^{-1}    \\
\hline             
\multicolumn{2}{c}{\footnotesize $^\dagger$ The 0.3\% factor represents the rescaling needed} \\
\multicolumn{2}{c}{\footnotesize  to compare with CHIME's field of view.} \\
\end{tabular}
\end{table}

We also want to compare our model's expected fluence distribution with CHIME's, both for the 115-burst sample and for two subpopulations of distant and close events, which in \citet{chime21} are separated at a DM threshold of 500~pc~cm$^{-3}$. 
The DM threshold separating the close and distant events was converted to a redshift threshold:~first, 
we subtracted the average ${\rm DM_{local}}$ contribution of our reference sample to obtain a ${{\rm DM_{lim}} = 375\mathrm{~pc~cm}^{-3} }$.~Then, from the function $P(z|\mathrm{DM_{\rm lim}})$ we derived the median redshift associated to it, 
$z_{\rm lim} = 0.315$.~We  
then used the same $10^3$ realizations randomly extracted above, now also considering the observed fluence of those bursts. Fig.~\ref{fig:fludist} shows the average cumulative distribution arising from those extractions for the two source subsamples, 
above and below $z_{\rm lim}$, indicated respectively by the red circles and squares, together with their 5th and 95th percentiles shaded areas. For the same model parameters as in Fig.~\ref{fig:npz} {(see Table~\ref{tab:simuparam})}, we calculated the expected fluence distributions for the two subsamples, and show them in Fig.~\ref{fig:fludist} as the blue solid curves: the general behaviour of the data is well reproduced, despite a small deficit at medium fluences, which could indicate a wider or flatter intrinsic energy distribution.
For comparison we also indicate the case with no magnetospheric absorption as the blue dashed curve.

\section{Discussion} \label{sec:summ}

In the gravitational self-lensing model the energy of 
bursts produced by single hotspots anchored to the magnetosphere of strongly magnetized NSs can be greatly
amplified along our line of sight when the burst occurs 
close to the caustic behind the star. The probability of burst amplification by lensing depends on the geometrical configuration of the system, mainly through 
the angular separation $i$ between the rotation axis and the caustic line, and the hotspot colatitude $\xi$.
If both angles are small, the NS rotation makes the hotspot spend most of its time close to the caustic line, thus enhancing the probability of its bursts being seen, while if both are large and comparable, there is a small, but non-zero, portion of its rotation which can be significantly amplified: the former sources appear then to us as often repeating, while the latter ones as one-off FRBs.  
This explains in a natural way the 
apparent dichotomy between FRB sources observed as repeaters and non-repeaters  
in terms of a single population of repeating systems. At the same time the model gives rise to a powerlaw-like distribution of high energy events as observed in 
the active repeaters investigated so far. Moreover it alleviates the energy requirements for individual bursts.

In this paper we carried out a detailed application of the GSL model to the burst energy distribution and behaviour of the best studied highly-active repeaters.
We first focused on \dod\ and showed that the observed double-peaked energy distribution of its FRBs can be readily explained and modelled by two antipodal rotating hotspots, giving rise to different components because of the different lensing of their bursts. 
In our interpretation one hotspot lies close to the caustic line behind the NS all along its rotation (the sum of $i$ and $\xi$ must be $\lesssim5$\textdegree) where large amplifications occur frequently, thus producing the higher energy component. 
In particular matching the highest observed FRB energies requires $|i-\xi| \lesssim0.1$\textdegree.
The second hotspot (which need not be precisely antipodal) is located in front of the NS, its bursts remain virtually unaffected by lensing and thus give rise to the lower-energy component. 
Therefore the observed log-normal distribution of the latter represents the seed distribution at the source, barring a slight redshift due to the proximity of the hotspot to the NS. It is similar in shape to the ones commonly observed from radio pulsars and RRATs, though at far higher energies \citep[see e.g.][]{Mickaliger2018, Shapiro-Albert2018}. 
 
The study of the burst energy distributions of repeaters  \ven, \ventidu\ and \dic\ produced similar results, in line with the model's prediction that most sources that are seen as highly repeating will be well-aligned systems with favourable geometry and some variation in the individual seed energy distributions. {In this scenario, for the four sources here studied the intrinsic total energies of the bursts are actually lower by a factor $3-30$ with respect to the observed ones.}

By extension the GSL model envisages the presence of a low and a high-energy component in many other repeating FRBs, as long as a second, nearly-antipodal hotspot is active on the close side of the NS, besides the strongly amplified hotspot lying near the caustic behind the NS. 
On the other hand if the hotspots are offset from the caustic line by $|i-\xi| \gtrsim 30$\textdegree\, only modest amplifications will be attained and the resulting energy distribution produced will be similar to the seed energy distribution of the bursts. Therefore the model predicts the existence of a larger population of repeaters, whose bursts remain virtually non-amplified.
Detecting such population as well as the lower component in the fluence distribution of individual repeaters displaying amplified bursts is challenging owing to the limiting sensitivity of present generation radio telescopes coupled to the rejection of low-DM, low-SNR events from population studies to avoid contamination from Galactic sources{, which would be indistinguishable from FRB sources that are close-by and within the same DM range} (see e.g. Appendix C in \paperone\ and \citealt{chime21}). 

The time evolution and long-term periodicity of \dod, whose bursts are seen to decrease in energy and then disappear completely, agree well with the predictions of the GSL model in the case of precession. In fact precession periodically modulates the inclination $i$ thus bringing the hotspot behind the NS further from the caustic and gradually decreasing the energy of the amplified events until a single low-energy distribution 
results from the merging of the components from the two hotspots.
Since in this scenario non-lensed events from this source would be observable at any time, the periodic disappearance of its FRBs requires the presence of an additional mechanism at play. 
One likely possibility is that the absorption of radio waves expected to occur 
inside the NS magnetosphere
prevents us from seeing bursts when their trajectory would cross the closed region of the magnetic field lines at large distance from the star, i.e., when the emission takes place quite far from the line of sight, be it in front or behind the star. 
This would also hamper the observability of the aforementioned population of non-amplified sources, as most of the systems with large values of $|i-\xi|$ would have their emission taking place in the region where the effect of magnetospheric absorption is dominant. {Furthermore, with such magnetospheric absorption, sources with hotspots that are not nearly antipodal could show either only amplified or only non-amplified bursts, as only one of the two circumferences traced by the hotspots might pass through the cones in which bursts are not absorbed by the magnetosphere.}

In the framework of the GSL model we presented also a simple application of the scenario envisioned for repeating sources to all FRBs. To derive the expected burst distribution with respect to both redshift and fluence, we assumed a population of sources tracking the NS formation history, each with randomly-oriented antipodal hotspots and 
log-normal distribution of seed energies of width comparable to that of the repeaters studied here, but with their mean values themselves log-normally distributed to account for intrinsic strength variability among sources. 
The FRB data consisted of a subsample of the first CHIME catalogue above its completeness threshold of 5~Jy~ms, and we also took into account possible biases due to selection effects which introduce uncertainty in the distance determination.
Comparisons with expected fluence and redshift distributions showed good agreement: we found degeneracies between the parameters of the seed energy distribution and also those of magnetospheric absorption along the burst signal trajectories (the mechanism we proposed in the case of highly-active repeaters), which could not be disentangled at present. {The number of bursts required per source, although highly variable depending on the shape of the intrinsic energy distribution assumed, remains lower than the one exhibited by very active repeaters such as \dod, \ven, or FRB~20240114A \citep[see e.g.][]{Zhang2025hyperactive}. This points towards the fact that, in the case of the log-normal energy distributions explored for the general population, either not all magnetars produce FRBs, or that the assumption of a uniform rate for all sources may be too simplistic, as already explored in previous studies \citep[e.g.][]{James2023}.}

{For simplicity, in both comparisons with the reference CHIME subsample only the uncertainties arising from the probabilistic association of redshift and DM were taken into account, while statistical uncertainties were neglected, as was the fact that fluence values in the CHIME intensity catalogue are only lower limits \citep[][]{chime21}. For the first catalogue, baseband data allowing a proper determination of the fluence was available for a small subset of bursts \citep[][]{Michilli2021,CHIME2024baseband}, which would have left very few bursts after the quality- and completeness-related cuts described in \citet{chime21} and Sect.~\ref{ssec:chime}. However, in the case of the second CHIME catalogue just released \citep[][]{CHIME2026second}, baseband data are expected to become available in the near future for roughly two thousand bursts (D. Michilli, private communication), which will allow an order of magnitude improvement in the number of bursts, with measured fluences for all of them.}

{As with all gravitational-lensing based scenarios, the GSL model envisions the production of two images of the same seed burst.\footnote{Barring sources so close to the star that for some angular separations only the first direct image is visible (see Appendix A in \paperone).} Detection of double-peaked events consistent with lensing would be further confirmation of the model, and searches for signatures of lensing have been already carried out on FRB data, although mostly focused on foreground lenses \citep[e.g.][]{Zhou2022,Kader2022}, finding a handful of potential candidates \citep[][]{Krochek2022,Chang2025,Xiong2026}.
For the parameters considered in this paper, $M_{NS} = 1.4 \, M_{\odot}$ and $R = 50\, R_g$, the time delay between the two images is $\Delta t \sim7 \, \upmu$s for bursts amplified by a factor 10, and it scales as $\Delta t \propto 1/a$ \citep[e.g.][]{Bakala2023}: since in the GSL model one-off sources are more likely to be observed as highly amplified bursts, these FRBs might have sub-microsecond delays, making detection impractical with current instrumentation \citep[for example, CHIME baseband data have a resolution of $2.56 \, \upmu$s; see however the techniques explored in][]{Kader2022}. The case of repeating FRBs is arguably more optimistic, since in the scenario discussed in Sect.~\ref{sec:rep} we expect to be also observing bursts with amplifications $\sim$$5-10$: provided the intrinsic duration of the peak is small enough, some cases of double images might be recovered from baseband data, although the identification will be made harder by the overlap with the effects of plasma lensing, scintillation and scattering \citep[e.g.][]{Cordes2017,Kader2022}.}



In its simple formulation developed so far, the GSL model shows a remarkable agreement with several of the key properties of both single FRB sources and the general population. Future work will present refined population synthesis through Monte Carlo methods, exploring multiple possibilities for the seed source properties and parameters, and letting us track the fractions of repeating and one-off sources as a function of distance, which could not be implemented in the current framework. 
The comparison of these synthetic populations with future larger samples of FRBs {such as the baseband-data second CHIME catalogue would provide an optimal test bench to break the model degeneracies and} accurately characterize the nature and intrinsic properties of the sources powering these events.

\vspace{-0.24cm}\section*{}
{\small 
This paper is dedicated to the memory of Pavel Bakala and Maurizio Falanga, dearly missed colleagues of ours who made important contributions to this line of research.\\
RLP would like to thank C.~W.~James for his support with the results of the \textsc{zdm} code and D.~Michilli for discussions on the baseband catalogue of CHIME events. We would like to thank the anonymous referee for the useful comments and suggestions about this paper. RLP acknowledges financial support from INAF's grants ``Uncovering the optical beat of the fastest magnetised neutron stars (FANS)'', ``PULSE-X'' (PI: Papitto),  ``Chasing the optical counterpart in FRBs and Magnetars with SiFAP2@TNG and SiFAP-SoFT@Cassini'', PI: Ambrosino, and both RLP and LS acknowledge financial support from the Italian Ministry of University and Research (MUR), PRIN 2020 (prot. 2020BRP57Z) ``Gravitational and Electromagnetic-wave Sources in the Universe with current and next-generation detectors (GEMS)''.~SD acknowledges financial support from the Grant FERMI LAT (Accordo ASI n. 2023-17-HH), and the support of the INAF -- Istituto di Radioastronomia di Bologna during the realization of this work. AP acknowledges  that part of the research activities described in this paper were carried out with contribution of the NextGenerationEU funds within the National Recovery and Resilience Plan (PNRR), Mission 4 -- Education and Research, Component 2 -- From Research to Business (M4C2), Investment Line 3.1 -- Strengthening and creation of Research Infrastructures, Project IR0000026 -- Next Generation Croce del Nord.~The visualization of data, sketches, and functions in this manuscript was carried out with \texttt{Wolfram Inc. Mathematica}, \texttt{MATLAB}, \texttt{gnuplot}, and \texttt{GeoGebra}.
}

\bibliography{bibo}{}
\bibliographystyle{aasjournal}

\end{document}